\documentclass[letterpaper, 10 pt, conference]{ieeeconf}  

\IEEEoverridecommandlockouts                              

\usepackage{graphicx} 

\title{\LARGE \bf
A Neural Network Based on Synchronized Pairs of Nano-Oscillators
}

\author{Damir Vodenicarevic$^{1}$, Nicolas Locatelli$^{1}$ and Damien Querlioz$^{1}$
\thanks{$^{1}$C2N, CNRS, Univ. Paris-Sud, Universit\'e Paris-Saclay, 91405 Orsay}%
}

\begin{document}

\maketitle
\thispagestyle{empty}
\pagestyle{empty}

\begin{abstract}

Artificial neural networks are intensively used to perform cognitive tasks such as image classification on traditional computers. With the end of CMOS scaling and increasing demand for efficient neural networks, alternative architectures implementing neural functions efficiently are being studied. This study leverages the demonstrated frequency tuning capabilities of compact nano-oscillators and their synchronization dynamics to implement a neuron using a pair of synchronized oscillators, and which features an unconventional response curve. We show that this compact neuron can naturally implement generic logic gates, including XOR. A simulated oscillator-based neural network is then shown to achieve results equivalent to standard approaches on two reference classification tasks. Finally, the performance of the system is evaluated in the presence of oscillator phase noise, an important issue of oscillating nanodevices. These results open the way for the design of alternative architectures adapted to efficient neural network execution.

\end{abstract}

\section{INTRODUCTION}

Artificial neural networks play an important role in natural data processing and cognitive tasks such as language translation, or  image classification. However, these networks are currently run by traditional computers or graphic cards using an important number of floating point operations to simulate essential neural non-linearities. 
With the increasing demand for high performance neural networks, alternative computational architectures capable of emulating neurons more efficiently are gaining considerable interest \cite{roy_brain-inspired_2014,vincent_spin-transfer_2015}.
Drawing inspiration from oscillatory behaviors observed in the brain,
multiple studies have proposed leveraging the complex dynamics of networks of coupled oscillators to achieve efficient neuro-inspired computation
\cite{nikonov_coupled-oscillator_2015,cotter_computational_2014,vodenicarevic_nanotechnology-ready_2017}. 
These approaches are gaining momentum with recent advances in nano-technology providing compact high frequency nano-oscillators with current or voltage-tunable natural frequencies, such as
superconducting Josephson junctions\cite{ovchinnikov_networks_2013},
spin-torque nano-oscillators \cite{locatelli_efficient_2015},
or oxide-based oscillators \cite{sharma_phase_2015}.

However, some proposed approaches rely on varying couplings between oscillators \cite{kaluza_computation_2013} which are difficult to implement. Others only implement binary threshold operations \cite{yogendra_coupled_2015}.
In this work, we propose an alternative approach and introduce a neuron design using a pair of coupled, synchronized oscillators. 
We derive the analytic expression of its activation function and 
show that this neuron is capable of representing generic logic gates, and assess its capacity to learn these gates through gradient descent. 
We then simulate the dynamics of artificial neural networks using these neurons on two standard machine learning tasks, and verify their robustness to the intrinsic noise of nano-oscillators. 

\section{Oscillator-Based Neuron}
   \begin{figure}[thpb]
      \centering
      \includegraphics[scale=0.95]{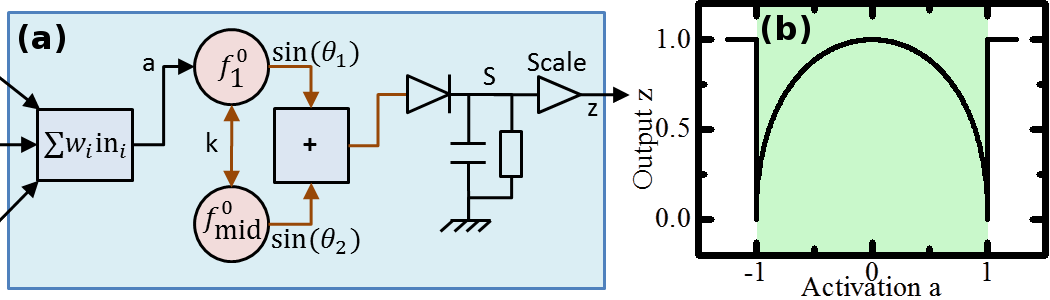}
      \caption{(a) Simplified schematic of the proposed neuron. Red circles are oscillators. (b) Neuron activation function for $f^0_\mathrm{amp}=2k$. The neuron operates in the phase locked (green filled) region.}
      \label{schema}
   \end{figure}
The proposed neuron circuit shown in Fig.~\ref{schema}(a) relies on a pair of bidirectionally coupled oscillators kept in a synchronized state. 
The Kuramoto model is a generic model that can predict the behavior of multiple types of oscillators \cite{acebron_kuramoto_2005,flovik_describing_2016}. We use it to model the two oscillators: 
\begin{equation}\label{kuramoto}
	\frac{\dot{\theta_1}}{2\pi}=  f^0_1 + k \sin\left(\theta_2-\theta_1  \right) \enskip;\enskip  \frac{\dot{\theta_2}}{2\pi}=  f^0_2 + k \sin\left(\theta_1-\theta_2  \right)
\end{equation}
where $\theta_i$ is the phase of oscillator $i$, $k$  the coupling constant  between between the nano-oscillators, typically obtained through electric or proximity coupling.
The natural frequency of the second oscillator is kept constant $f^0_2=f^0_\mathrm{mid}$, while the natural frequency of the first oscillator is tuned by the weighted sum of the neuron inputs $a=\sum_i w_i \mathrm{in}_i$ as follows: $f^0_1=  f^0_\mathrm{mid} + a f^0_\mathrm{amp} $.
A peak detector is applied to the sum of the two oscillator signals $\sin(\theta_1) + \sin(\theta_2)$. 
For simplicity, in the rest of the paper, the result of peak detection $S$ is rescaled between 0 and 1 and represents the output $z= (S-\sqrt{2})/(2-\sqrt{2})$.

The output, or activation function, of the neuron shown in Fig.~\ref{schema}(b) can be expressed analytically. The peak detection yields $S=2\cos\left( \frac{\theta_1-\theta_2}{2} \right)$ which gives, with the frequency locking synchronization condition $\dot{\theta_1}=\dot{\theta_2}$ applied to (\ref{kuramoto}):
\begin{equation}\label{Seq}
S=\sqrt{2+2\sqrt{1-\left(\frac{a f^0_\mathrm{amp}}{2k}\right)^2} } \enskip ; \enskip z=\frac{S-\sqrt{2}}{2-\sqrt{2}}
\end{equation}
The paraboloid-shaped activation function (\ref{Seq}) is non-monotonous, unlike the sigmoid-like activation functions used conventionally in artificial neural networks.

To assess the capabilities of this unconventional neuron, we learn two-input logic gates by simulating (\ref{Seq}) using realistic parameters $f^0_\mathrm{mid}=500\mathrm{MHz}$, $f^0_\mathrm{amp}=10\mathrm{MHz}$, $k=6\mathrm{MHz}$ typical of spin-torque nano-oscillators \cite{chen_spin-torque_2016}, an exponential-decay peak detector, and the standard gradient descent learning algorithm. Two input weights, and one bias value are tuned, and the learning is run for $1,000,000$ different sets of initial weight and bias values to assess the robustness of the results.
\begin{table}[h]
\caption{Proportion of states converging to  correct logic function.}
\label{tablogic}
\begin{center}
\begin{tabular}{|c|c|c|c|}
\hline
Target gate & Oscillator neuron & Sigmoid neuron & Threshold neuron   \\
\hline
XOR& 15.8 & 0 & 0  \\
XNOR& 29.8 & 0 & 0  \\
Others & 100 & 100 & 100 \\
\hline
\end{tabular}
\end{center}
\end{table}

Table \ref{tablogic} shows the percentage of initial conditions leading to successful logic gate learning, and compares it with a traditional sigmoid neuron, and with a threshold neuron. The results show that the oscillator neuron successfully learns all the gates the threshold or sigmoid neurons can learn, and with favorable initial conditions it also learns to solve XOR and XNOR, which is not achievable by a sigmoid nor a threshold. This is allowed by the non-monotonous activation function of the oscillator pair: each of its lobes ($a<0$ or $a>0$) can be used as classical non-linear activation functions while using different lobes for different input examples allows more advanced operations. The energy landscapes of XOR and XNOR exhibit multiple and different local minima which accounts for their non-perfect and different scores. As the amplitude of the gradient with oscillator neurons is higher than with sigmoid neurons, the learning rate parameter of the gradient descent is adjusted to obtain comparable learning iteration numbers.

\section{Application for Machine Learning Tasks}

In this section we simulate more comprehensive neural networks using our oscillator-based neuron.
First, a simple 1-layer neural network containing 3 neurons is trained by gradient descent (or equivalently the perceptron learning rule for the threshold neurons) on a standard task: the Iris classification data set.

   \begin{figure}[ht]
      \centering
      \includegraphics[scale=0.95]{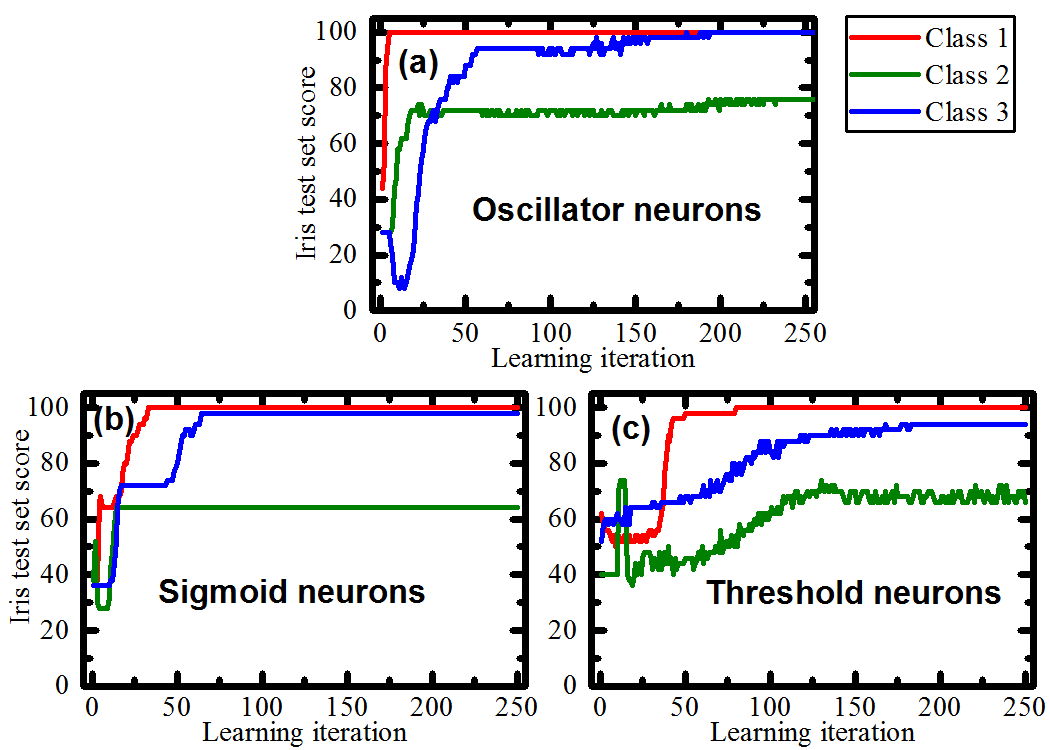}
      \caption{Iris test set classification rate evolutions during learning for the 3 classes using (a) oscillator-based neurons, (b) classical sigmoid neurons (c) threshold neurons.}
      \label{iris}
   \end{figure}

The evolutions of the test set classification rates of the 3 classes during learning are plotted in Fig.~\ref{iris}(a), and compared to a sigmoid layer (Fig.~\ref{iris}(b))  and to a threshold layer (Fig.~\ref{iris}(c)). 
The results show that oscillator-based neurons converge to similar recognition rates for classes 1 ($100\%$), 3 ($\geq 95\%$) and to slightly increased recognition rates for the heavily non-linearly separable class 2 ($76\%$ using the oscillator neuron, $\geq 65\%$ using a sigmoid or a threshold neuron). 
This result confirms that the proposed neuron successfully operates in a neural network.


We then train a more complex two-layer (300 hidden, 10 output) neural network to classify the handwritten digits of the MNIST database, using the back-propagation algorithm. 
The  final test set classification rate of $95.7\%$ is on par with classical sigmoid networks ($95.3\%$, \cite{lecun_gradient-based_1998}), which validates the effectiveness of the proposed neuron in multi-layer neural networks.

\begin{figure}[h]
      \centering
      \includegraphics[scale=1.0]{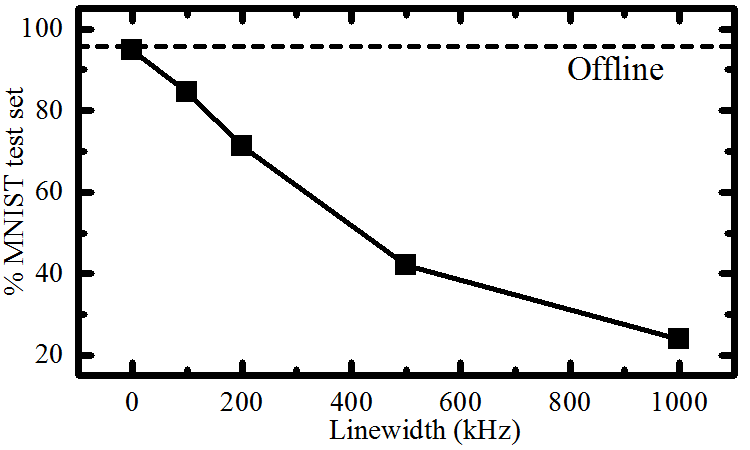}
      \caption{MNIST test set classification rate as a function of the oscillator linewidth.}
      \label{mnist}
   \end{figure}

Finally, Fig.~\ref{mnist} assesses the effects of oscillator noise by showing the MNIST test set classification rate of the fully simulated noisy oscillator system, as a function of the oscillator linewidth. The classification rate stays over $84\%$ for a linewidth of $100\mathrm{kHz}$, a noise value typical of spin-torque nano-oscillators \cite{tsunegi_high_2014}, which suggests that the proposed approach could be implemented using such nano-devices.
\section{Conclusion}
We proposed a compact neuron implementation using a pair of synchronized oscillators and validated, through simulation, its use for machine learning. Our results show that the proposed neuron is a valid functional alternative to sigmoid or threshold neurons, and naturally computes its parabola-shaped transfer function through its intrinsic dynamics.

This work opens the way towards hardware neural network accelerators exploiting the intrinsic dynamics of modern nano-devices. Further studies are still required, as the learning schemes rely on external computers while exploiting nano-device physics for the learning itself would be more attractive.
\clearpage

\addtolength{\textheight}{-12cm}   





\bibliographystyle{IEEEtran}

\end{document}